# Observation of a metastable intermediate during solid-solid phase transformation in response to rapid compression

Michael R. Armstrong[1*], Harry B. Radousky[1], Ryan A. Austin[1], Elissaios Stavrou[1], Hongxiang Zong[2], Graeme J. Ackland[2], Shaughnessy Brown[3], Jonathan C. Crowhurst[1], Arianna E. Gleason[4], Eduardo Granados[3], Paulius Grivickas[1], Nicholas Holtgrewe[5], Hae Ja Lee[3], Tian T. Li[1,] Sergey Lobanov[5], Joseph T. McKeown[1], Robert Nagler[3], Inhyuk Nam[3], Art J. Nelson[1], Vitali Prakapenka[6], Clemens Prescher[7], John D. Roehling[1], Nick E. Teslich[1], Peter Walter[3], Alexander F. Goncharov[5] and Jonathan L. Belof[1]

[1]Lawrence Livermore National Laboratory, Livermore CA.
[2]University of Edinburgh, Scotland, UK.
[3]SLAC National Accelerator Laboratory, Menlo Park, CA
[4]Los Alamos National Laboratory, Los Alamos, NM
[5]Geophysical Laboratory, Carnegie Institution of Washington, Washington DC.
[6]Argonne National Laboratory, Argonne, IL.
[7]University of Cologne, Cologne, Germany.

*Corresponding author: armstrong30@llnl.gov

**Abstract**

In order to probe the mechanism of solid-solid phase transformations, we have applied ultrafast shock wave compression (120 picosecond duration) and ultrashort (130 femtosecond) x-ray diffraction at the Linac Coherent Light Source (LCLS) to probe the compression-induced phase transition pathway in zirconium. Surprisingly, rather than transform from α-Zr to the more disordered hex-3 equilibrium ω-Zr phase, in its place we find the formation of a non-equilibrium body-centered cubic (bcc) metastable intermediate. Theoretically hypothesized for several decades, this bcc intermediate state has now been found to be dynamically stabilized under uniaxial loading at sub-nanosecond timescales. Molecular dynamics simulations of shock-wave propagation in zirconium predict this transformation via the dynamical intermediate state. In contrast with longer timescale experiments where the phase diagram alone is an adequate predictor of the crystalline structure of a material, our recent study highlights the importance of metastability and time-dependence in the kinetics of phase transformation at extreme conditions.

**Summary:** In place of a more disordered phase, we observe the formation of a non-equilibrium metastable bcc phase along an otherwise known phase transformation pathway, highlighting the importance of time-dependence in solid-solid phase transitions at extreme conditions.

First-order phase transformations proceed from nucleation and growth mechanisms out of non-equilibrium metastable states toward a state of equilibrium, with solid-solid transitions presenting particular complexity owing to the multiple (atomistic) pathways and deformation mechanisms potentially involved. (*1–14*) Through the use of a laser ablation induced shock wave to drive a rapid change in volume, and subsequent *in situ* examination *via* x-ray diffraction using the Linac Coherent Light Source (LCLS), we have studied the non-equilibrium phase transition process pertaining to the α→ω structural rearrangement in zirconium (Zr) at the earliest stages of nucleation.(*3, 15*) While this solid-solid phase transition has been well studied at larger length and time scales, the extant observations of over-pressurization needed to overcome its relatively slow kinetics on the microsecond timescale – with the transformation not observed until a pressure of nearly 10 GPa above the phase boundary – and relation to proposed pathways have remained a challenge to experimentalist and theorist alike.(*4, 5, 7, 8*) By focusing in detail on the very short (sub-nanosecond) times preceding and during the transition, we find that the mechanism of transformation occurs through a body-centered cubic (bcc) non-equilibrium intermediate state, in stark contrast to prior experimental findings at longer (diamond anvil cell and gas-gun) timescales. Furthermore, the ω-Zr phase is never observed at any pressure (from ambient to shock melting at 130 GPa) on the sub-nanosecond timescale, but rather the anomalous appearance of the bcc structure is consistently formed in its place, suggesting the need for new hypotheses that shed light on the transformation mechanism. The surprising observation of the formation of a high-symmetry non-equilibrium phase, rather than the more disordered hex-3 ω-Zr equilibrium phase, highlights the importance of time-dependence in understanding the behavior of matter at extreme conditions.

Dynamic compression experiments, which obtain thermodynamic state data under extreme conditions of pressure and temperature, typically require compression time scales of at least 10 nanoseconds to ensure equilibration of the final compressed state(*16*). Our shorter time scale (~100 ps) compression experiments have allowed us to observe intermediate states along the material transformation path, analogous to transition states observed in time-resolved characterization of chemical reactions.(*17, 18*) These fast compression experiments employed uniaxial compression, which accesses far-from-equilibrium mechanical states under extreme shear stress(*19*). Conventional picosecond time-scale compression experiments(*19, 20*) can, by characterizing the flow of material, determine the time-dependent bulk stress/strain state within a sample. Such experiments have been used to observe a wide range of anomalous behavior under ultrafast compression, including extreme elastic deformation(*19, 21, 22*) (related to a lack of plastic deformation on this time scale(*21, 23–25*)), simultaneous phase transformation and plasticity(*26*), with high throughput.(*27*)[*] While these experiments have provided information about the bulk response of materials, they did not have access to femtosecond x-ray diffraction, and therefore did not provide a direct measure of the atomic structure.

---

[*]Further comments about ultrafast shock experiments, elastic-plastic response, and their connection to longer time scale experiments are in the supplemental information.

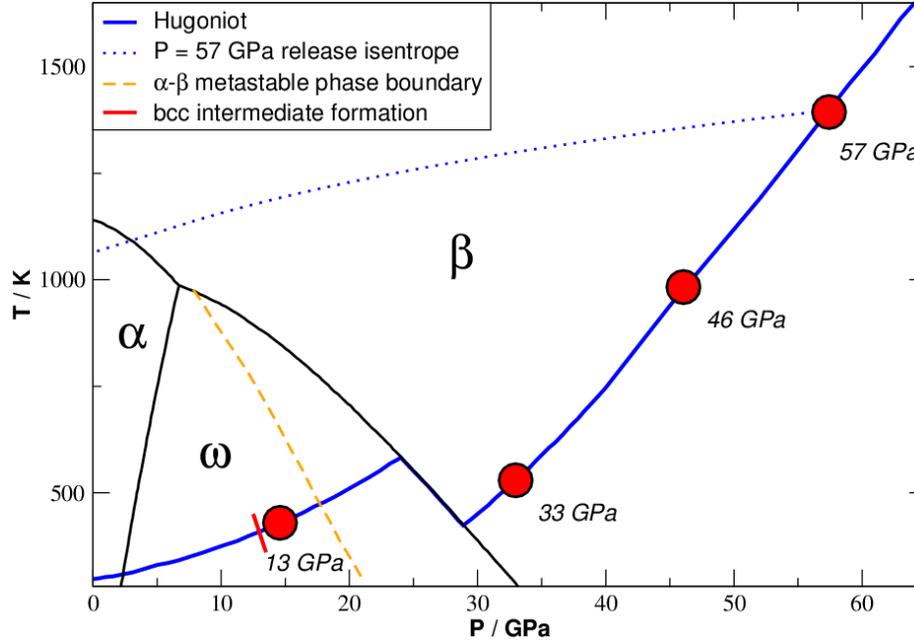

Figure 1: The phase diagram of Zr with the shock Hugoniot and the α, β, and ω phase boundaries as reported by Greeff [36] and which reflect aggregate gas-gun and diamond anvil cell data from multiple sources. The four shock wave experiments conducted in this study consist of final Rankine-Hugoniot pressures of 13, 33, 46 and 57 GPa as determined from a combination of x-ray diffraction analysis and impedance matching to the adjacent Al standard. The release isentropes for all experiments (that from the P = 57 GPa experiment is shown) have a final state outside of the β-Zr field of stability and thus β would not be expected to form upon release. The metastable phase boundary between the hcp α and bcc β phases have been calculated from metastable branches of Greeff's multiphase equation of state and define the region where, if the hex-3 ω phase were absent (or kinetically hindered) the transition from α directly to β would be favored by the bulk Gibbs free energy difference. The dynamically stabilized bcc intermediate state, formed along the α-ω transformation pathway (but never reaching the thermodynamic end-state of the ω phase on the sub-nanosecond timescale in this experiment) is demarcated near 13 GPa, which is the lowest pressure where it has been observed in this study.

With the advent of femtosecond time resolution x-ray diffraction at the Linac Coherent Light Source (LCLS), it has now become feasible to explore the transient states of materials under extremely rapid compression with nm/ps resolution. (*24, 28–31*) By combining the 100 ps laser drive available at the MEC beamline of LCLS with the 100 fs x-ray capability, we can begin to address some the of fundamental questions concerning phase transformations: Are intermediate states ordered or disordered at the atomic scale? If ordered, what are the intermediate states? How does rapidly applied shear stress influence the transition pathway? Answers to these questions have the potential to greatly enhance material synthesis by providing more sophisticated tools to control material transformations beyond the currently available adjustment of thermodynamic variables such as temperature and pressure.

At longer (greater than nanosecond) timescales, the dynamic behavior of Zr and its various solid-solid phase transitions have been relatively well-studied and found to exhibit complex phase transition behavior.(*2–5, 7, 8, 10–13, 15, 32–35*) The Zr phase diagram, shock Hugoniot and the highest pressure release isentrope thermodynamic paths followed in this study are depicted in Fig 1. In addition to the equilibrium stability fields, the metastable phase boundary between the α and β phases has been calculated from metastable extensions of the Gibbs free energy models of Greeff (*36*) and plotted. The significance of the α-β metastable phase line pertains to the stability of the bcc phase at pressures *lower* than the ω-β phase boundary, as found in the experimental results of this study.

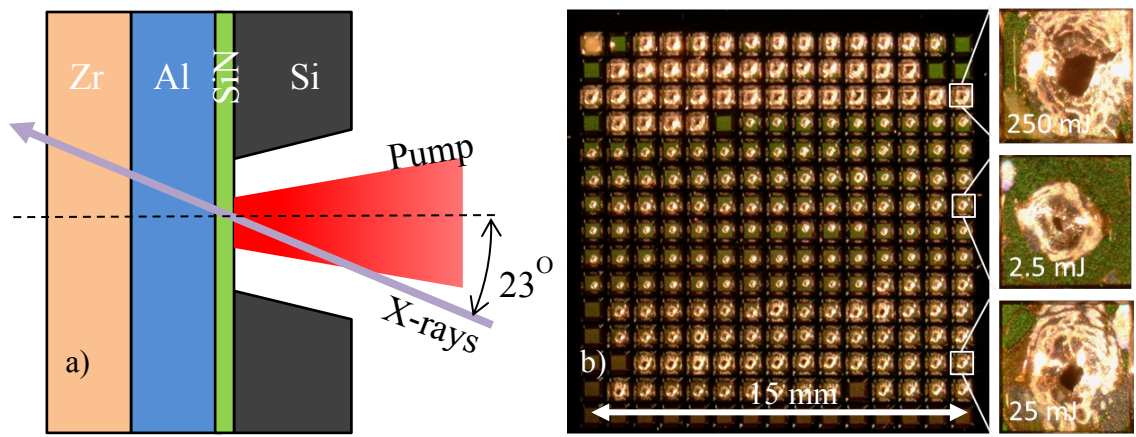

**Figure 2: Laser driven shock compression experiments with XRD at LCLS: a) schematic of the experiment; b) the pump side of the sample after the experiments with close-ups of sample sites where shots were taken at 3 different pump energies.**

Experiment description. A schematic of the experiment and pictures of the sample are shown in Figure 2. The x-ray wavelength was 1.301 angstroms. A 2.2 µm Al ablator (coated on a thin SiN window) is driven by a ~120 ps FWHM duration laser pulse with a fast (sub-10 ps, consistent with previous work - see Armstrong et al.(*19*)) initial rise, launching a shock wave of similar ~120 ps FWHM duration into the ablator. The intensity contrast of the pulse is sufficient to avoid significant preheating in the sample, as demonstrated from estimates of Al peak shifts. Swinburne et al.(*30*) applied a similar duration (170 ps) pulse with no fast rise, but which is close enough in drive conditions for qualitative comparisons in the discussion section. The shock wave transits the ablator and enters the 1.7 µm thick Zr sample, where rapid compression initiates a phase transition to one of several possible final states (shown in Fig. 1) depending on the drive energy. After a variable delay, a ~100 fs duration x-ray pulse is used to obtain an x-ray diffraction (XRD) pattern from the sample, providing structural information during the compression wave transit through the sample. Using this method, the progress of phase transformations in the Zr sample can be tracked with better than picosecond time resolution.

Data was taken at five laser drive energies: 2.5, 25, 30, 50 and 250 mJ. This resulted in peak pressures ranging from 10-100 GPa in the Al, corresponding to 13-130 GPa in the Zr, which is obtained by shock impedance matching with the Al ablator.(*37*) Full sets of diffraction patterns for the 4 lower energies are shown in Figures S5-S8 in the supplemental information. A complete discussion of the highest energy shot (250 mJ), which involved melting and refreezing of the Zr, will be given separately.(*38*) Selected diffraction patterns for drive energies 2.5, 25 and 50 mJ are shown in Figure 3. Hydrodynamics simulations designed to emulate the conditions of each corresponding experiment (shown in Figure 3) illustrate the progress of the compression wave through the Al/Zr sample as a function of time. For the 2.5, 30 and 50 mJ data, the peak pressure in the Al ablator was obtained by estimating the peak strain from the diffraction patterns (as illustrated in Figures S9-S11) and using the known Hugoniot of Al. This yielded values of ~10, 36 and 44 GPa, respectively. From shock wave impedance matching, this corresponds to values of ~13, 46 and 57 GPa in the Zr. All peak pressures obtained for both the Al and Zr are shown in Table T1 in the supplemental information.

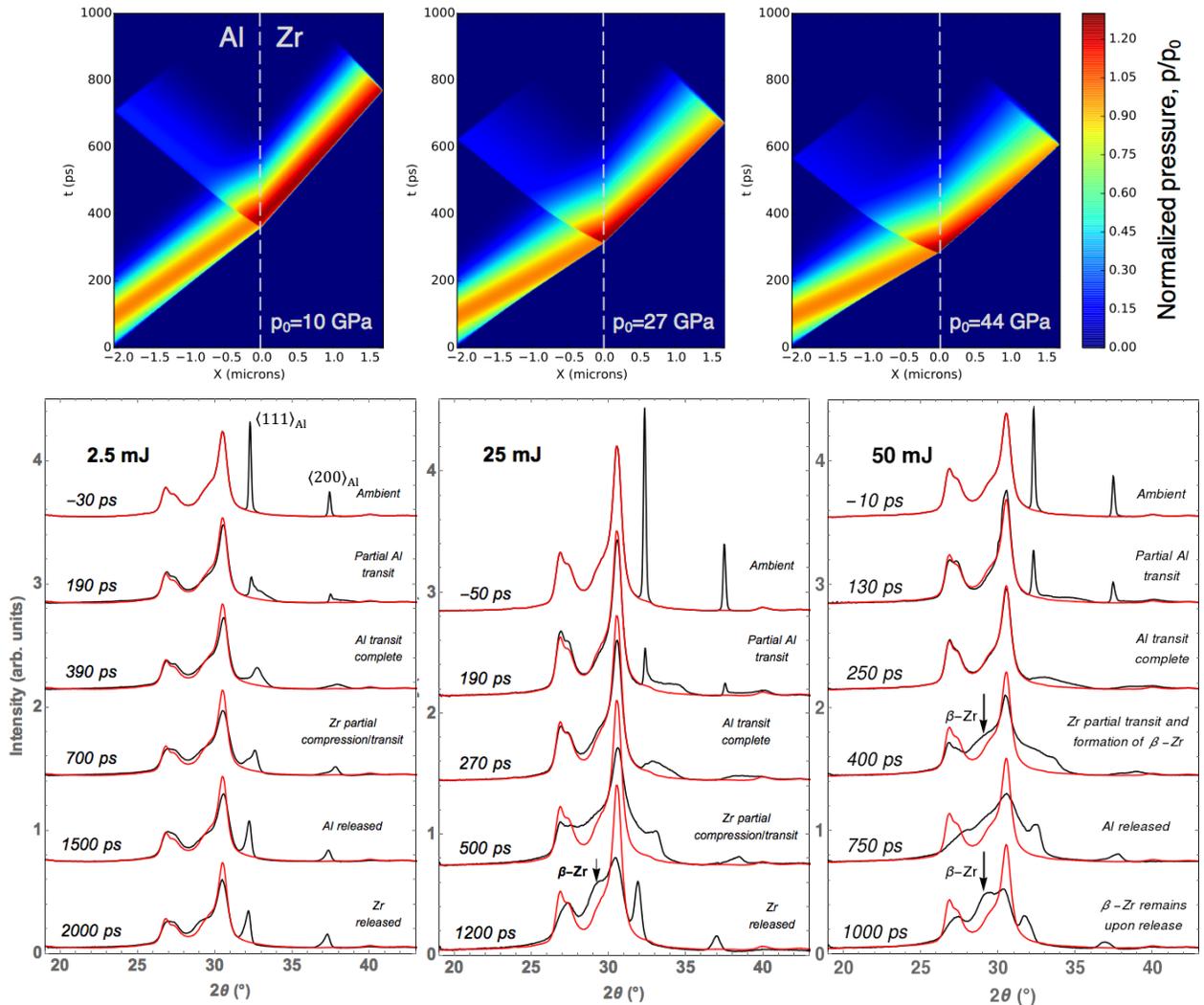

**Figure 3:** Experimental data (bottom) and hydrodynamics simulations (top) for three different drive energies and several time delays between the drive-pulse and the x-ray probe pulse, showing correspondence in timing between the simulations and the data. The drive pulse is incident from the left in the simulations. Red traces are the Zr background with Al diffraction peaks removed for comparison to time-resolved data in black. The simulations illustrate wave propagation via Lagrangian X-t diagrams of pressure, where a pressure boundary condition is applied to the left-hand surface of the Al layer as a proxy for the laser drive. The peak pressure in the Al ablator at the leftmost drive surface, $p_0$, is given for each drive energy. Data at intermediate times are shown in Figures S5-S8.

For the 25 mJ data, a more detailed analysis was applied, as described in the supplemental information. In this case, a series of hydrodynamic simulations with different pulse shapes and peak pressures gave simulated XRD patterns which were compared to the experimental diffraction patterns, and the best match for the data (examples are shown in the Figure S4 of the supplemental information) was chosen. This gave a maximum pressure of 27 GPa in the Al which corresponds to a pressure of 33 GPa in the Zr, and also provided insight into the time dependence of the compression profile. The peak pressure for both analysis methods was given via the Al EOS by the peak strain (see Figs. S9-11) – we conservatively estimate the error in pressure to be less than

20% for the 13 GPa data, and less than that for higher pressure data. All data gave final pressure estimates well within the equilibrium stability range of the relevant phase.

Prior to shock wave arrival, ambient Al peaks were not temperature shifted to lower angles (i.e. larger d-spacings and lower density) after the laser drive pulse arrival for any drive energy, indicating that remaining unshocked Al was not significantly heated by the drive pulse prior to shock arrival. For all data, ambient Al peaks entirely vanished on a time scale consistent with compression wave propagation through the Al ablation layer. Subsequent to release, diffraction indicated solid Al for all drive energies, with peaks shifted to lower angles consistent with high temperature solid phase of Al at zero pressure.

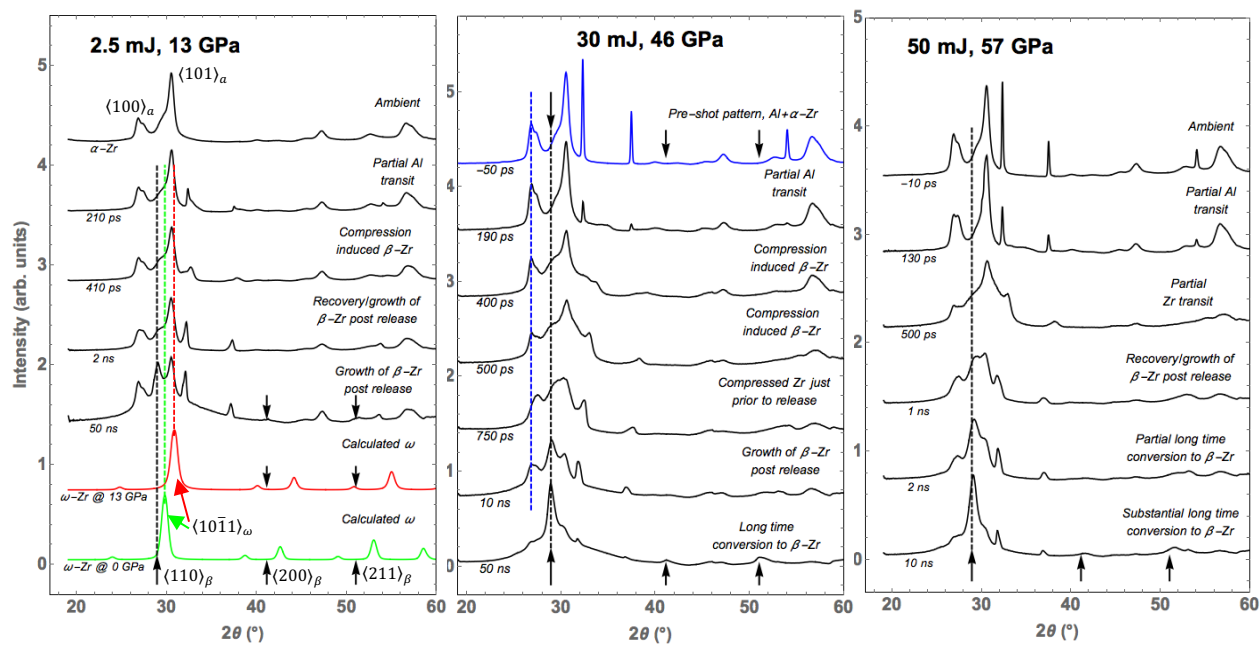

**Figure 4: Diffraction data at three drive energies/Zr peak pressures showing the appearance of β-Zr under compression and subsequent to release.** Calculated ω-Zr patterns at 0 (green) and 13 (red) GPa are shown with the 2.5 mJ data, with green and red dashed lines showing the position of the $\langle 10\bar{1}1 \rangle$ ω-Zr peak cited by Swinburne et al. Black arrows and black dashed lines show positions of β-Zr peaks, and the blue dashed line shows the position of the $\langle 100 \rangle$ α-Zr peak. Black arrows are placed to show both the presence of β-Zr (e.g. 2.5 mJ, 50 ns) and the absence of β-Zr peaks in ω- and α-Zr patterns (e.g. ω-Zr at 0 GPa).

Structural Phase Transitions. Over the conditions obtained by this experiment, ω-Zr is never observed. Calculated ω-Zr patterns (using POWDER CELL(*39*) - details in the supplemental information) are shown in Figure 4 (left panel) for shots at 2.5 mJ corresponding to 0 and 13 GPa peak pressures. A compression-induced peak near 29º first appears subsequent to compression in the Zr, but prior to release. This peak forms to the right of the expected position of the $\langle 110 \rangle$ β-Zr peak, but to the left of the expected (at ambient pressure) $\langle 10\bar{1}1 \rangle$ ω-Zr peaks,(*40*) consistent with the formation of β-Zr under compression. Compression pressure-shifts all α-Zr peaks in the 25-32° range to higher angles with consistent timing for the arrival of the compression wave in the Zr sample. Higher order α-Zr peaks were observed to shift (indicating compression) and move back to their ambient locations (indicating release), respectively for 2.5 and 25-50 mJ data, consistent with the timing of transit through the Zr layer at the given pressure (see Figure 4, and S5-S8).

We observe nucleation of the bcc β-Zr phase under shock compression in the Zr at the 25, 30 and 50 mJ drive energies (corresponding to equilibrium peak pressures of 33, 46 and 57 GPa respectively), within ~100 ps subsequent to shock arrival in the Zr layer (see Figures 4, and S6-8 in the supplemental information). An example of the appearance of this phase is shown in Figure 4 for the 30 mJ drive energy, where the appearance a β-Zr peak is evident near 29° in the pattern at 400 ps, approximately 100 ps after shock arrival in the Zr sample. After the shock wave has propagated through the entire sample, but not yet released (at 750 ps), β-Zr remains evident while higher-angle shifted α-Zr peaks (corresponding to higher than ambient density) indicate the sample is under compression. In the 2.5 mJ data, β-Zr is observed upon compression at 410 ps. For all data, a growing phase is observed subsequent to release of pressure in Zr at long time scales, whose peak positions are consistent with β-Zr rather than ω-Zr.

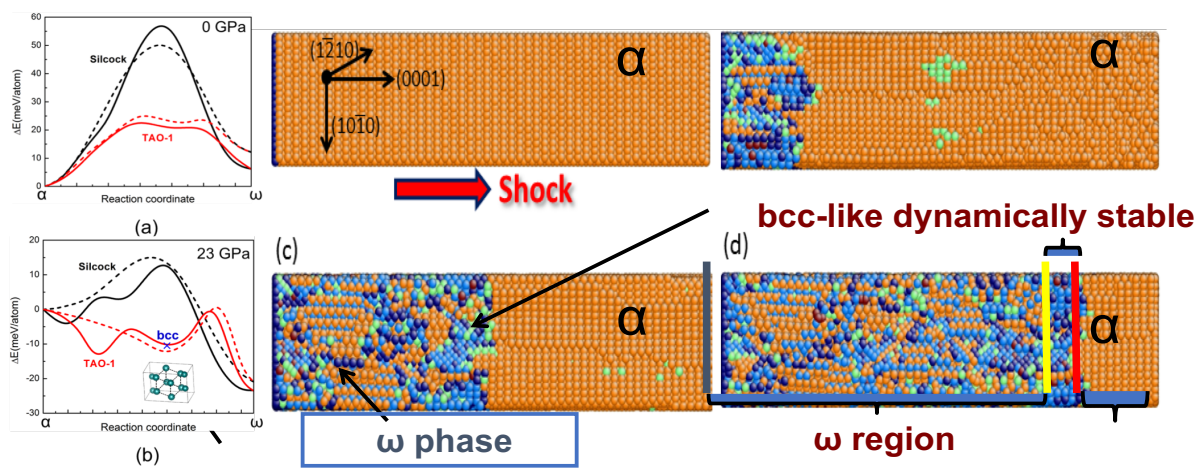

**Figure 5:** Static transition path energetics comparing TAO-1 (red) and Silcock (black) paths calculated with DFT (solid lines) and the potential (dashed line) at (a) ambient and (b) 23 GPa under hydrostatic conditions (c,d) snapshots from the shock MD (uniaxial compression along the ⟨0001⟩ direction) with hcp (orange), bcc (blue) and omega (blue and orange stripes) corresponding to a shock pressure of 13 GPa. The leading edge of the shock wave generates a dynamically stable bcc intermediate. Please see Fig S19 of the supporting information which demonstrates even larger amounts of bcc intermediate being generated in shocks with a peak pressure of 15 GPa and for the ⟨10$\bar{1}$0⟩ and ⟨1$\bar{2}$10⟩ orientations.

In parallel with the experiment, an interatomic potential for Zr was developed to reproduce the equilibrium phase diagram.(*41*) With molecular dynamics (MD) simulations of bulk compressive transformation (*42, 43*) the α–ω and α–β transformations occur at the expected temperature and pressure conditions. However, when MD simulations of shock waves (piston velocity = 0.48 and 0.54 km/s resulting in a shock stress of 13 and 23 GPa, respectively) were performed, it was found that the α-Zr transformed into a metastable bcc phase for several picoseconds before undergoing a second transformation to the expected ω-Zr phase. Tracking the orientation relation from the MD simulations suggested the TAO-1 transformation pathway proposed by Trinkle *et al.* based upon density function theory (DFT) evaluations of several proposed mechanisms(*9*). The energetics of the TAO-1 pathway given by our interatomic potential reproduces well the pathway given by the more accurate density functional theory. Both our potential and DFT calculations give an intermediate bcc state at an energy minimum along the TAO-1 path, separated from ω-Zr by an energy barrier of more than 10 meV/atom. Furthermore, the amount of dynamically generated bcc-Zr was greatly increased when shock loading was applied to the ⟨10$\bar{1}$0⟩ and ⟨1$\bar{2}$10⟩ crystallographic orientations (as shown Figure S16 of the Supporting Information). While the MD simulations show an admixture of both metastable bcc and ω-Zr, depending upon the orientation the amount of bcc can be as high as 50 %; this result is consistent with the x-ray diffraction patterns observed on the non-textured Zr samples. This atomistic study explains the observation that bcc-Zr is rapidly formed behind the shock wave and survives in a metastable fashion.

Referring again to the phase diagram shown in Fig. 1, we propose the following mechanistic pathway as suggested by the experimental data and simulations. As the α-Zr is rapidly compressed along the Rayleigh line under uniaxial shock wave loading the system evolves toward the lowest free energy state, which is ω-Zr (for pressures less than 25 GPa), but for which a bcc transition state is a local free energy minimum; in order for the transition to proceed the Zr lattice must remain in the transient bcc state for at least some finite period of time until the energy barrier for final lattice rearrangement to hex-3 can be overcome. The kinetic residence time for dwelling in the bcc intermediate is apparently on the order of a 100 picoseconds or so. During that period of time, the pressure is further increased until the shock wave has become steady, reaching the Rankine-Hugoniot end state. By the time that the kinetic residence would have ended, the system pressure now exceeds the 18 GPa threshold of the metastable α-β transition and therefore, with the ω-Zr phase kinetically hindered, the "next best" phase to thermodynamically exist is that of β-Zr. Eventually, of course, the hex-3 ω-Zr phase will form but not in the time-frame of the sub-nanosecond experiments conducted in this study.

Discussion.

We do not observe ω-Zr as a final state, despite pressures that are well within the equilibrium stability field of ω-Zr at 13 GPa, nor as an intermediate in higher energy shots which extend up to peak pressures of 130 GPa in the Zr. Our observation of an ordered, bcc Zr intermediate along the transformation path confirms the hypothesis of Usikov *et al.*(*7*) that an ordered intermediate phase exists between the initial α-Zr and final ω-Zr phases. This long-lived intermediate state is likely the origin of the remarkable metastability of α-Zr and the exceedingly slow kinetics of the α→ω Zr phase transition which has remained a mystery on longer compression timescales.(*4, 5, 7, 8*) Growth of the new phase (subsequent to its nucleation under compression) is obtained after stress release on greater than 1 ns time scales, suggesting that further growth of the new phase is obtained via the relaxation of shear stress subsequent to strong uniaxial compression.

Our observations are contrary to the conclusion of Swinburne *et al.*(*30*) (see further comments in the supplemental information), who assign a final ω-Zr state subsequent to compression under similar drive conditions (170 ps duration drive), sample geometry (2 μm thick Zr), and estimated pressures (7.6 and 22.1 GPa for intermediate and high pressure shots). Diffraction patterns obtained in our experiment (Figures 3 and 4, 2.5 mJ) are similar to patterns shown in Fig. 2b of Swinburne *et al.*, yet we obtain data over a wider angular range (which provides more peaks for identification of the new phase) and note a quantitative discrepancy between the expected ω-Zr peak positions and the observed peak positions, which are consistent with β-Zr. Over the wide angular range of our experiment, the appearance of peaks at ~41° and ~51° is well correlated (shown with arrows in Figure 4) with known peaks of β–Zr, but not ω-Zr. Further, a comparison of our observed diffraction patterns to calculated ω-Zr diffraction patterns at the compression (13 GPa) and ambient (0 GPa) pressures shows no evidence of the ω phase (also see Figure 4 and Figure S14 in the supplemental information), although our experiment obtains a pressure in the range of pressures where Swinburne *et al*. claim to see ω-Zr.

The absence of ω-Zr under compression in our experiment suggests that transformation to ω-Zr is kinetically limited over a ~100 ps compression time scale. The transition from α-Zr to ω-Zr would be expected to occur between approximately 5 GPa and 30 GPa under equilibrium conditions (see Figure 1). However, the observed pressure where the α–ω transition occurs can be considerably higher due to kinetics under dynamic compression, ranging from 3 GPa under quasi-static conditions, up to 11 GPa for ns scale compression.(*32*)

For samples with comparable hafnium and oxygen content as used in this experiment (see figure S1 in the supplemental information), the α-ω phase transition is not observed in velocimetry measurements in gas gun-driven shock compression experiments at pressures less than 10 GPa (*32, 33*). To within the sensitivity of the experiment, our data also indicate that the α-ω phase transition does not complete over the time scale of our experiment, which suggests that kinetics limit the complete transition to ω-Zr. Yet, partial conversion to a bcc phase of Zr is evident in the 2.5 mJ data (corresponding to 13 GPa peak pressure in Zr), suggesting that bcc Zr observed here is a metastable state along the shock compression path to ω-Zr. This short time-scale experimental observation is consistent with longer time scale impact driven compression experiments which (post-experiment) recover β-Zr(*4*) and β-Ti(*5*) (Ti has a phase diagram similar to Zr) in a thin layer near the sample drive surface, static work on Ti-V(*6*), and theoretical work(*7–9*) – see the supplemental information for more details.

The absence of the ω-Zr phase at the sub-nanosecond time scale is notable, since it has always been seen in longer time experiments (to the appropriate final pressure), and is stable enough to be present in recovery experiments.(*10, 34*) Here, our observations suggest that rapid compression formed the bcc Zr intermediate on a ~100 ps time scale, followed by the release of pressure prior to the formation of the final ω-Zr state, leaving the bcc β-Zr as a metastable phase (over the time scale of the experiment - no sample was recovered afterwards). Assuming quasi-isentropic release from shocked states (a release from 57 GPa is shown in Fig. 1), Zr shocked to pressures less than ~60 GPa should revert to α-Zr, but on very long scales all data exhibited growth of β-Zr subsequent to the release of pressure. Although the growth of β-Zr may be thermally driven at zero pressure at temperatures above 1143 K (see Fig. 1), this explanation is not consistent with our observations. In particular, thermal expansion shifts of α-Zr peaks in >5 ns data (i.e. subsequent to release) are well below what would be expected from α-Zr at or above 1143 K (See Fig. S13). Further, at late times the Al ablator remains solid, indicating an Al temperature of less than ~950 K, which (assuming thermal contact between the Al ablator and the Zr sample) indicates that Zr is at less than this temperature as well.

Based on the experimental data, molecular dynamics and analysis derived from classical nucleation theory (see supporting information), we hypothesize that the reason we do not observe ω-Zr (but rather see an abundance of β-Zr) on the several hundred picosecond timescales being probed here is due to the formation of the bcc β-Zr transition intermediate along the phase transformation pathway toward (but never reaching) ω-Zr. At this increased stress, α-Zr is unstable with respect to β-Zr and quickly transforms. However, a significant free energy barrier still exists between bcc and ω-Zr and there is not sufficient time for this slower process to occur within the 100-300 picosecond time interval over which compression is maintained in this experiment. On longer compression timescales found in experiments utilizing gas-gun(*35*), pulsed power(*32*) and diamond anvil cells(*13*), the relatively slow kinetics of the α→ω Zr phase transition has long been unexplained(38,41); through interrogating the transformation at sub-nanosecond timescales with a bright x-ray source, our observations suggest that a bcc β-Zr long-lived intermediate state is likely the origin of this behavior.

Our results demonstrate an alternate mechanism for transition to β-Zr under the conditions of this experiment. It is due to relaxation of shear stress subsequent to the rapid, strong uniaxial shock compression, seeded by nuclei of β-Zr formed under compression. This is clearly demonstrated in our molecular dynamics simulations, which shows the shock-driven transformation proceeding along a pathway which includes β-Zr as an intermediate state. We showed that this intermediate β-Zr state, which is unstable under ambient conditions, is a deep local minimum at 23 GPa, stable relative to α-Zr with no kinetic barrier, but separated from the ω-Zr ground state by a large energy barrier.

## Acknowledgments

Lawrence Livermore National Laboratory is operated by Lawrence Livermore National Security, LLC, for the U.S. Department of Energy, National Nuclear Security Administration under Contract DE-AC52-07NA27344  We acknowledge J.M. Zaug for helpful discussions and planning support.  M.R.A., H.B.R, R.A.A., E.S., J.C.C, P.G., T.T.L., J.T.M., A.J.N, J.D.R., N.E.T. and J.L.B. gratefully acknowledge the LLNL LDRD program for funding support of this project under 16-ERD-037. GJA and HZ thank EPSRC and ERC for funding.

**Author Contributions –**
**M.R.A. and H.B.R. wrote the manuscript, with contributions from R.A.A., J.L.B., E.S. and P.G.  The experiments were planned by M.R.A., H.B.R., J.L.B. H.J.L. and A.F.G. LCLS MEC data was taken by M.R.A., H.B.R., S.B., A.E.G., E.G., P.G., N.H., H.J.L., S.L., B.N., I.N., V.P., C.P., P.W. and A.F.G. Sample preparation was performed by J.C.C, P.G., T.L., and J.D.R. Sample characterization was performed by P.G, A.J.N., J.T.M. and NET.  Data analysis was performed by M.R.A., H.B.R., R.A.A., J.L.B., E.S., J.C.C., P.G., C.P. and A.F.G.  Molecular dynamics calculations were performed and analyzed by HZ and GJA. Hydrodynamic simulations were performed by M.R.A., R.A.A. and J.L.B.**